\documentclass[draft,showkeys,showpacs,eqsecnum,nofootinbib]{revtex4}
\renewcommand{\theequation}{\arabic{equation}}
\def\beq{\begin{equation}}
\def\eeq{\end{equation}}
\def\bea{\begin{eqnarray}}
\def\eea{\end{eqnarray}}\def\nn{\nonumber}
\def\pr{\prime}
\def\na{\nabla}
\def\pa{\partial}

\def\nn{\nonumber}

\begin{document}
\title{Global embeddings and hydrodynamic properties of Kerr black hole}
\author{Soon-Tae Hong}
\email{soonhong@ewha.ac.kr} \affiliation{Department of Science
Education and Research Institute for Basic Sciences, Ewha Womans
University, Seoul 120-750, Republic of Korea}
%\date{June 14, 2008}
\date{\today}
\begin{abstract}
In the presence of a rotating Kerr black hole, we investigate hydrodynamics of the massive particles and
massless photons, to construct relations among number density, pressure and internal energy
density of the massive particles and photons around the rotating Kerr black hole and to study an accretion onto 
the black hole. On equatorial plane of the Kerr black hole, we investigate the bound orbits of the massive 
particles and photons around the black hole to produce their radial, azimuthal and precession 
frequencies. With these frequencies we study the black holes GRO~J1655-40 and 4U~1543-47, 
to explicitly obtain the radial, azimuthal and precession frequencies of the massive particles in the 
flow of perfect fluid. We next consider the massive particles in the stable circular orbit of radius of $1.0~ly$ 
around the supernovas SN 1979C, SN 1987A and SN 2213-1745 
in the Kerr curved spacetime, and around the potential supermassive Schwarzschild black holes M87, NGC 3115, NGC 4594, 
NGC 3377, NGC 4258, M31, M32 and Galatic center, to estimate their radial and azimuthal frequencies, 
which are shown to be the same results as those in no precession motion. 
The photon unstable orbit is also discussed in terms of the impact parameter of the photon trajectory. 
Finally, on the equatorial plane of the Kerr black hole, we construct
the global flat embedding structures possessing (9+3) dimensionalities outside and 
inside the event horizon of the rotating Kerr black hole. Moreover, on the plane we investigate the warp 
products of the Kerr spacetime.
\end{abstract}
\pacs{02.40.Ky, 04.20.Dw, 04.40.Nr, 04.70.Bw, 95.30.Lz}
\keywords{accretion; Kerr black hole; hydrodynamics; bound orbits; GRO~J1655-40; 4U~1543-47; warp product; global embedding}
\maketitle

%%%%%%%%%%%%%%%%%%%%%%%%%%%%%%%%%%%%%%%%%%%%%%%%%%%%%%%%%%%%%%%%%%%%%%%%
\section{Introduction}
\setcounter{equation}{0}
\renewcommand{\theequation}{\arabic{section}.\arabic{equation}}
%%%%%%%%%%%%%%%%%%%%%%%%%%%%%%%%%%%%%%%%%%%%%%%%%%%%%%%%%%%%%%%%%%%%%%%%

An accretion onto black holes has attracted
lots of attentions from physicists and astronomers. As for the Schwarzschild black hole~\cite{sch}, 
luminosity coming from interstellar gas accretion of the 
black hole were investigated by exploiting spherically symmetric 
stead-state accretion on a nonrotating black hole at rest~\cite{shapiro73}. This model 
was later generalized~\cite{shapiro74} to more realistic case of a rotating 
Kerr bleck hole~\cite{kerr63}. According to the accretion onto the rotating black hole, the luminosity has been shown 
to increase as the angular momentum of the black hole increases. 
For the values of $a=0$ and 0.998 $M$ of the Kerr black holes with $a$ and $M$ being the spin parameter and the mass 
of the black hole, the X-ray spectrum produced by the accretion 
disk was calculated to show that the equatorial observer sees a much harder spectrum at high energies than 
the polar observer does and this effect is more 
definite for the black holes of larger angular momentum~\cite{cunningham75}. 
The general relativistic hydrodynamic simulations 
of the accretion were also performed to investigate the cases where the fluid has a finite angular momentum in 
the Kerr curved manifold~\cite{bambi10}. 
The particle flux originated from collisions 
of weakly interacting particles was also constructed near the Kerr black hole~\cite{silk11}.

Phenomenologically, the luminosity has been studied to investigate the 
supernovas. Specifically, the supernova SN 1987A 
was interpreted in terms of its progenitor star nature and 
its interaction with circumstellar matter~\cite{chevalier87}. The supernova SN 1987A was also shown to be 
a blue supergiant, rather than a red supergiant, possessing a mass 20 $M_{\odot}$~\cite{arnett89}. The neutrino 
mass limits were calculated by exploiting the neutrino signals from SN 1987A~\cite{rosner87}. Moreover, 
the velocity structure of a ring expanding at 10.3 $km~sec^{-1}$ was studied around 
the supernova SN 1987A~\cite{crotts91}, and the hard X-rays from the remnant of the supernova SN 1987A 
was also reported in the band containing lines from the decay of $^{44}$Ti~\cite{gre12}. Based on the works on the peak luminosities 
of the supernova~\cite{kasen}, the X-ray luminosity was found 
to be manifestly constant over more than one decade years 
in the supernova 1979C~\cite{nasa}. This observation of the high and steady luminosity has been considered 
as plausible evidence for a stellar-mass (around $(5-10)~M_{\odot}$) black hole accreting material from 
a supernova fallback disk. It was shown that, due 
to pair formation, stars heavier than ($20-30)~M_{\odot}$ become dynamically unstable~\cite{barkat67}. The nucleosynthesis 
of helium cores with ($64-133)~M_{\odot}$ was studied to conclude that it corresponds to main-sequence star masses around 
($140-260)~M_{\odot}$~\cite{heger02}. The supernova SN 2213$-$1745 was detected to be superluminous and 
to have an estimated progenitor mass of ($100-250)~M_{\odot}$~\cite{cooke12}. 
On the other hand, lots of progresses have been made in global embeddings of various
black holes with coordinate singularities~\cite{deser97,hong00,hong001,hong002,hong003}. 
As novel ways of removing the coordinate singularities, the higher dimensional 
global embedding Minkowski space (GEMS) of the black hole solutions are
subjects of great interest both to mathematicians and to
physicists. In differential geometry, four dimensional Schwarzschild metric is shown not 
to be embedded in ${\mathbf R}^{5}$~\cite{spivak75}. Recently, (5+1) dimensional 
GEMS structure for the Schwarzschild black hole has been consructed~\cite{deser97} to
study a thermal Hawking effect on a curved manifold~\cite{hawk75} associated with an Unruh effect~\cite{unr} in the higher dimensional spacetime. 

On the other hand, progress on the thin-disk oscillations has been made in terms of the dispersion relations~\cite{okazaki,kato90,silbergleit,kato01}. 
They studied 
axially symmetric, isothermal pulsations of a geometrically thin disk rotating around a nonrotating relativistic 
object~\cite{okazaki}. According to this model, there exists a characteristic that 
global pulsation modes are trapped in an inner region of the disk. They have also described the wave trapping in the radial direction by 
exploiting the dispersion relation. Later, one-armed corrugation waves were shown to be 
trapped in the innermost region of accretion disk by introducing the dispersion relation~\cite{kato90}. The modes of the oscillation trapped within the inner region of the accretion disk were also investigated to focus on the nearly incompressible perturbations of the inner disk~\cite{silbergleit}. The oscillations on thin disk were revisited to study the characteristics of the
disk oscillations and their excitation mechanisms, together with dispersion relation~\cite{kato01}. It is noteworthy to observe that, to treat the particle motions moving around the relativistic compact objects, all the above authors have used the dispersion relations  
rather than the effective potential of the relativistic compact object.

In the realistic astrophysical situation, we will assume in this paper that the
supernovas such as SN1987A and SN 1979C are decaying into the rotating Kerr 
black hole and the particles consist of
the hydrogen and helium gas almost fully ionized to perform thermal vibration, as well as gravitational 
one in a relativistic effective potential. 
The gravitational mode constrained by the fluid dynamic equations can
generate the radio wave envelope, especially in the vicinity of
a black hole, of the high frequency thermal radiation, which
play a major role in producing the circular ring with sensible
luminosity around the black hole~\cite{potter09}. 

In this paper, we will exploit the rotating relativistic object and its relativistic effective 
potential to investigate the accretion onto the black hole. Specifically, we will study the bound orbits of the massive 
particles and photons around the Kerr black hole to produce their radial, azimuthal and precession 
frequencies, on the equatorial plane of the Kerr black hole. Moreover, on this plane we will investigate 
the mathematical aspects of the Kerr spacetime such as the warp products and the GEMS.

Now, the rotating Kerr black hole embedded in an asymptotically flat spacetime
is described by the four-metric~\cite{kerr63,wald}
\beq
ds^{2}=-\frac{\Delta-a^{2}\sin^{2}\theta}{\Sigma}dt^{2}
       -\frac{4Mar\sin^{2}\theta}{\Sigma}dt d\phi\
       +\frac{A\sin^{2}\theta}{\Sigma}d\phi^{2}
       +\frac{\Sigma}{\Delta}dr^{2}+\Sigma d\theta^{2},
\label{kerrmetric} \eeq
where 
\bea
\Delta&=&r^{2}-2Mr+a^{2},\nn\\
\Sigma&=&r^{2}+a^{2}\cos^{2}\theta,\nn\\
A&=&(r^{2}+a^{2})^{2}-\Delta a^{2}\sin^{2}\theta,
\eea
and $M$ and $a$ ($0\le a\le M$) are the mass and angular
momentum per unit mass of the black hole, respectively.
Here one can easily see that in the limit of $a=0$ the above Kerr
metric reduces to the Schwarzschild case. We consider the four
velocity given by $u^{a}=dx^{a}/d\tau$
where one can choose $\tau$ to be the proper time
(affine parameter) for timelike (null) geodesics.

%%%%%%%%%%%%%%%%%%%%%%%%%%%%%%%%%%%%%%%%%%%%%%%%%%%%%%%%%%%%%%%%%%%%%%%%
\section{Hydrodynamics of massive particles and photons on Kerr black hole}
\setcounter{equation}{0}
\renewcommand{\theequation}{\arabic{section}.\arabic{equation}}
%%%%%%%%%%%%%%%%%%%%%%%%%%%%%%%%%%%%%%%%%%%%%%%%%%%%%%%%%%%%%%%%%%%%%%%%

%%%%%%%%%%%%%%%%%%%%%%%%%%%%%%%%%%%%%%%%%%%%%%%%%%%%%%%%%%%%%%%%%%%%%%%%
\subsection{Set up of hydrodynamical massive particles on Kerr black hole}
%\setcounter{equation}{0}
%\renewcommand{\theequation}{\arabic{section}.\arabic{equation}}
%%%%%%%%%%%%%%%%%%%%%%%%%%%%%%%%%%%%%%%%%%%%%%%%%%%%%%%%%%%%%%%%%%%%%%%%

In this subsection, we briefly recapitulate the hydrodynamics of massive particles which was 
studied mainly in~\cite{shapiro74}. To do this, we first study the fundamental 
equations of relativistic fluid dynamics which can
be obtained from the conservation of particle number and
energy-momentum fluxes.  In order to derive an equation for the
conservation of particle numbers one can use the particle flux
four-vector $nu^{a}$ to yield 
\beq
\nabla_{a}(nu^{a})=\frac{1}{\sqrt{-g}}\pa_{a}(\sqrt{-g}~ nu^{a})=0,
\eeq
where $n$ is the proper number density of
particles measured in the rest frame of the fluid of massive particles and
photons and $\nabla_{a}$ is the covariant derivative in
the Kerr curved manifold of interest and $g={\rm det}~g_{ab}$. For
steady state axisymmetric flow of the perfect fluid of the massive
particles and photons, the conservation of
energy-momentum fluxes is similarly described by the Einstein
equation 
\beq
\nabla_{b}T_{a}^{b}=\frac{1}{\sqrt{-g}}\pa_{b}(\sqrt{-g}~T_{a}^{b})=0,
\label{nbtab}
\eeq
where the stress-energy tensor
$T^{ab}$ for perfect fluid is given by 
\beq
T^{ab}=\rho
u^{a}u^{b}+P(g^{ab}+u^{a}u^{b}),
\eeq
with $\rho$
and $P$ being the proper energy densities and pressures of the
massive particles or photons, respectively. For the steady state axisymmetric flow of the perfect fluid of the
massive particles and photons, the equations $\nabla_{a}(nu^{a})=0$ and $\nabla_{b}T_{a}^{b}=0$
yield
\beq 4\pi
r^{2}nu^{r}\left(1+\frac{a^{2}}{r^{2}}\cos^{2}\theta\right)\sin\theta=C_{0},~~~
(P+\rho)r^{2}u_{i}u^{r}\left(1+\frac{a^{2}}{r^{2}}\cos^{2}\theta\right)\sin\theta=C_{i},
\label{flow2}
\eeq
where $C_{0}$ is the accretion rate of the massive particles,
and $C_{i}$ ($i=t,\phi$) are the other constants of the motion which can be
evaluated at infinity to yield the ratio $C_{\phi}/C_{t}=u_{\phi}/u_{t}=0$.  Exploiting the equations in 
(\ref{flow2}), one can derive the relations~\cite{shapiro74}
\begin{widetext}
\beq \frac{(P+\rho)^{2}}{n^{2}}
\frac{1-2Mr^{-1}+a^{2}r^{-2}+(1+a^{2}r^{-2}\cos^{2}\theta)u^{r}u^{r}}
{1+a^{2}r^{-2}[1+2Mr^{-1}\sin^{2}\theta
/(1+a^{2}r^{-2}\cos^{2}\theta)]}
=\frac{(P_{\infty}+\rho_{\infty})^{2}}{n_{\infty}^{2}}, \label{prho}
\eeq\end{widetext} where $n_{\infty}$, $P_{\infty}$ and
$\rho_{\infty}$ are the particle number density, pressure and internal energy
density of the fluid of the massive particles at infinity, respectively.

%%%%%%%%%%%%%%%%%%%%%%%%%%%%%%%%%%%%%%%%%%%%%%%%%%%%%%%%%%%%%%%%%%%%%%%%
\subsection{Photon hydrodynamics on Kerr black hole in extended formalism}
%\setcounter{equation}{0}
%\renewcommand{\theequation}{\arabic{section}.\arabic{equation}}
%%%%%%%%%%%%%%%%%%%%%%%%%%%%%%%%%%%%%%%%%%%%%%%%%%%%%%%%%%%%%%%%%%%%%%%%

In this subsection, we first extend the massive particle result (\ref{prho}) in the previous subsection to 
the generalized one by including the photon degrees of freedom in the presence of the Kerr black hole. 
We next construct the other hydrodynamic identities describing the rotating Kerr black hole case and the photon 
dynamics. (See (\ref{euler2}) and (\ref{radial}) below.) These newly constructed two identities are the extended ones 
of the previous work~\cite{shapiro73} which was studied only
on the case of the massive particles around the Schwarzschild black hole. 

We now construct dynamic relations relevant to both the massive particles and photons. To do this, 
here we introduce a new parameter 
$\kappa$ defined as 
\beq
\kappa=-g_{ab}u^{a}u^{b}
\eeq
which is one for timelike geodesics and zero for null 
geodesics. From now on, to figure out the massless photon dynamics we just insert $\kappa=0$ in the equations below. 
For the steady state axisymmetric flow of the perfect fluid of the photons, the equations 
\beq
\nabla_{a}(nu^{a})=0
\eeq
and (\ref{nbtab}) produce (\ref{flow2}).
Using the equations in (\ref{flow2}), one can derive the dynamic relations for the massive particles and/or photons
\begin{widetext}
\beq \frac{(P+\rho)^{2}}{n^{2}}
\frac{\kappa(1-2Mr^{-1}+a^{2}r^{-2})+(1+a^{2}r^{-2}\cos^{2}\theta)u^{r}u^{r}}
{1+a^{2}r^{-2}[1+2Mr^{-1}\sin^{2}\theta
/(1+a^{2}r^{-2}\cos^{2}\theta)]}
=\frac{(P_{\infty}+\rho_{\infty})^{2}}{n_{\infty}^{2}}[\kappa+(1-\kappa)u^{r}_{\infty}u^{r}_{\infty}], \label{prho2}
\eeq\end{widetext} where $n_{\infty}$, $P_{\infty}$ and $\rho_{\infty}$ are the particle number density, pressure and internal energy
density of the fluid of the massive particles and/or photons at infinity, respectively. Here, we observe 
that the massive particles initially at rest and the photons with an initial velocity $u_{\infty}^{r}\approx 1$, respectively,
at infinity travel toward the black hole along a conical surface of constant $\theta=\theta_{\infty}$ where
$\theta_{\infty}$ is the polar angle at infinity.  

Next, using the projection operators in (\ref{nbtab}) one can
obtain the general relativistic equation on the direction
perpendicular to the four-velocity~\cite{wald} 
\beq
(P+\rho)u^{b}\nabla_{b}u_{a}+(g_{ab}+u_{a}u_{b})\nabla^{b}P=0,
\eeq 
from which, after some algebra, we obtain the
radial component of the above equation for the steady state
axisymmetric accretion of the massive particles and/or photons on the rotating Kerr black hole of mass $M$
%\begin{widetext}
\bea &&\frac{1}{2}\frac{d}{dr}(u^{r}u^{r})+\kappa\frac{M(1-a^{2}r^{-2}\cos^{2}\theta)}{r^{2}(1+a^{2}r^{-2}\cos^{2}\theta)^{2}}
+\frac{Ma^{2}\sin^{2}\theta}{r^{4}(1+a^{2}r^{-2}\cos^{2}\theta)(1-2Mr^{-1}+a^{2}r^{-2})}\left(\frac{r}{M}
+\frac{B}{A}\right)u^{r}u^{r}
\nonumber\\
&&+\kappa\frac{Ma^{2}\sin^{2}\theta}{r^{4}(1+a^{2}r^{-2}\cos^{2}\theta)^{2}}\frac{B}{A}
+\frac{1}{P+\rho}\left(u^{r}u^{r}+\frac{1-2Mr^{-1}+a^{2}r^{-2}}{1+a^{2}r^{-2}\cos^{2}\theta}\right)
\frac{dP}{dr}=0.
\label{euler2} \eea
%\end{widetext}
Here we have introduced the function 
\beq
B=(r^{2}-a^{2}\cos^{2}\theta)\Delta-4Mr^{3},
\eeq
which is a
characteristic for the rotating Kerr black hole. We observe that, in (\ref{euler2}) as in the
other rotating Kerr black hole equations (\ref{flow2}) and (\ref{prho2}), the axisymmetric
gravitational terms due to the angular momentum of the black hole
vary as $a^{2}/r^{2}\leq M^{2}/r^{2}$, while the
terms independent of the black hole rotation vary as $M/r$. The Einstein equation in (\ref{nbtab}) can be
readily rewritten in another covariant form 
\beq
u_{a}\nabla_{b}((P+\rho)u^{b})+(P+\rho)u^{b}\nabla_{b}u_{a}+\nabla_{a}P=0.
\eeq
Multiplying this equation by $u^{a}$ and using 
the continuity equation $\nabla_{a}(nu^{a})=0$, one can project it on the
direction of the four-velocity to obtain 
\beq
nu^{a}\left(\nabla_{a}\frac{P+\rho}{n}-\frac{1}{n}\nabla_{a}P\right)=0,
\eeq
whose radial component yields 
\beq
\kappa\frac{d\rho}{dr}-\kappa\frac{P+\rho}{n}\frac{dn}{dr}+(\kappa-1)\frac{d P}{dr}
=\frac{\Lambda-\Gamma}{u^{r}}.
\label{radial} \eeq Here the energy loss $\Lambda$ and the energy
gain $\Gamma$ are introduced to set the decrease in the entropy of
the inflowing massive particles and/or photons equal to difference 
$\Lambda-\Gamma$. Here one notes that there exists no $a$-dependence in (\ref{radial}) even though 
this equation holds both for the static black hole and for the rotating one. 
In the limit of $\kappa=1$ and $a=0$, the results in (\ref{euler2}) 
and (\ref{radial}) reduce to the massive particle cases of the Schwarzschild black 
hole~\cite{shapiro73}.  Finally, we inform you that the equations (\ref{prho2}) and (\ref{euler2}) will be used in (\ref{hydrodynamics22}) and (\ref{hydrodynamics2}) of Section III 
to treat the hydrodynamics of the massive particles in the perfect fluid trapped in the circular orbit around the 
rotating relativistic compact object. 

%%%%%%%%%%%%%%%%%%%%%%%%%%%%%%%%%%%%%%%%%%%%%%%%%%%%%%%%
\section{Phenomenology of Kerr black hole}
\setcounter{equation}{0}
\renewcommand{\theequation}{\arabic{section}.\arabic{equation}}
%%%%%%%%%%%%%%%%%%%%%%%%%%%%%%%%%%%%%%%%%%%%%%%%%%%%%%%%

In order to investigate the phenomenological aspects of the Kerr black hole, we 
construct the frequencies of the massive particles along the azimuthal and 
radial directions. Exploiting these frequency formulas, we study the 
characteristics of the rotating black holes and supernovas with given masses. To this end,
we consider the equatorial plane on $\theta=\pi/2$ where we use 
the three-metric obtainable from the Kerr metric (\ref{kerrmetric}). 
In the Kerr metric (\ref{kerrmetric}), we have two Killing vector fields $\xi_{i}^{a}~(i=t, \phi)$
satisfying the Killing equations 
\beq
\pounds_{\xi_{i}}g_{ab}=\na_{a}\xi_{ib}+\na_{b}\xi_{ia}=0.
\eeq 
Using the above Killing equations and geodesic ones, one can readily obtain  
\beq
g_{ab}\xi_{t}^{a}u^{b}=-\epsilon,~~~g_{ab}\xi_{\phi}^{a}u^{b}=l,
\eeq
corresponding to the Killing vector fields, where $\epsilon$ and $l$ are 
the conserved energy per unit mass and angular momentum per unit mass, for the massive particles.
For the photons, we note that $\hbar \epsilon$ and $\hbar l$ are the total energy and
the angular momentum of the photons, respectively.

Now, we proceed to calculate the radial equation for the particle
on the equatorial plane whose metric is given by the three metric to yield
\beq
\frac{1}{2}u^{r}u^{r}+V(r)=0,
\eeq
where the effective potential is given by~\cite{wald}
\beq
V(r)=-\frac{\kappa
M}{r}+\frac{l^{2}}{2r^{2}}+\frac{1}{2}(\kappa-\epsilon^{2})\left(1+\frac{a^{2}}{r^{2}}\right)
-\frac{M}{r^{3}}(l-a\epsilon)^{2}.
\label{veff}\eeq Here one notes that the first and second terms
are the Newtonian and centrifugal barrier terms, respectively, and
the others are the general relativistic corrections including the
black hole rotating effects with the spin parameter $a$. The effective potential (\ref{veff}) now should fulfill the
condition 
\beq
\frac{dV}{dr}(r=r_{s,us})=0,
\eeq
from which we evaluate  
the radii $r_{s,us}$ of the stable and unstable bound orbits on the equatorial
plane for a given $\epsilon$ and $l$, to yield 
\beq
r_{s,us}=\frac{1}{2M}[l^{2}+(1-\epsilon^{2})a^{2}\pm
D^{1/2}],
\eeq
 where the upper (lower) sign refers to stable
(unstable) orbit and
\beq
D=[l^{2}+(1-\epsilon^{2})a^{2}-2\sqrt{3}M(l-a\epsilon)]
[l^{2}+(1-\epsilon^{2})a^{2}+2\sqrt{3}M(l-a\epsilon)].
\eeq
In oder to guarantee the positive value of $D$, $l$ should
satisfy the constraints that
\bea
l&\ge&M+(\sqrt{3}M+a-a\epsilon)^{1/2}(\sqrt{3}M-a-a\epsilon)^{1/2},~~~{\rm for}~a(1+\epsilon)\le\sqrt{3}M,\nn\\
l&\ge& -\sqrt{3}M+(\sqrt{3}M+a+a\epsilon)^{1/2}(\sqrt{3}M-a+a\epsilon)^{1/2},~~~{\rm for}~a(1+\epsilon)\ge\sqrt{3}M, 
\eea
respectively. In the limit of $a=0$, one
readily notes that the above $r_{s,us}$ reduces to
the well known Schwarzschild case for
$l\ge 2\sqrt{3}M$~\cite{wald}. 

The energy per unit mass of the particle in the circular orbit of
the radius $r=r_{s}$ is just the value of the effective potential
$V$ at that radius $V(r=r_{s})=0$, which, together with 
$\frac{dV}{dr}(r=r_{s})=0$, yields the energy $\epsilon_{c}$ per unit mass for a circular orbit
and the angular momentum $l_{c}$ per unit mass whose explicit forms are given in~\cite{bardeen72}. 
The angular frequency in the azimuthal direction for the zero mode of
the circular orbit with $\epsilon_{c}$ and $l_{c}$ at the stable bound orbit of $r=r_{s}$ is
then given by
\beq \omega_{\phi}=\frac{1}{r_{s}^{2}+a^{2}-2Mr_{s}}
\left[\frac{2Ma}{r_{s}}\epsilon_{c}+\left(1-\frac{2M}{r_{s}}\right)l_{c}\right].\label{omegaphi}\eeq
On the other hand, in the radial direction we have the harmonic motion frequency of the massive particles
of the form
\beq \omega_{r}=\frac{1}{r_{s}^{2}}
\left[2Mr_{s}-(l_{c}^{2}+(1-\epsilon_{c}^{2})a^{2})\right]^{1/2}.
\label{omegar}
\eeq
Here one notes that, since there exists some difference between $\omega_{\phi}$ and $\omega_{r}$ in general, the
massive particles perform the precession motion with the frequency 
\beq
\omega_{p}=\omega_{\phi}-\omega_{r}.
\eeq 
In examples with the particle angular momentum per unit
mass $l_{c}=5.0~M=2.211\times 10^{6}~(M/M_{\odot})~km^{2}~sec^{-1}$, one can
obtain the frequencies $(\omega_{\phi},\omega_{r})=(2034,1796)~(M/M_{\odot})~sec^{-1}$,
$(2014,1787)~(M/M_{\odot})~sec^{-1}$ and $(1994,1778)~(M/M_{\odot})~sec^{-1}$
to yield the precession rate $\omega_{p}=238~(M/M_{\odot})~sec^{-1}$, $227~(M/M_{\odot})~sec^{-1}$ and
$216~(M/M_{\odot})~sec^{-1}$, for the cases of $a=0.65~M$, $0.75~M$ and $0.85~M$, respectively. 
Next, in order to figure out their phenomenological aspects, 
we consider the black holes GRO~J1655-40 and 4U~1543-47 with given masses $M$ and spin parameters $a$: 
$6.30~M_{\odot}$ and $(0.65-0.75)~M$, and $9.40~M_{\odot}$ and $(0.75-0.85)~M$, respectively~\cite{shafee06}.
With these experimental data for the black hole masses and spin parameters, we obtain the explicit 
azimuthal, radial and precession frequencies corresponding to these black holes, as listed in Table I. In the above calculations we have used the formulas of 
$\epsilon_{c}$ and $l_{c}$ given in~\cite{bardeen72}.

%------------ Table 1 --------------------------------------------
\begin{table*}
%\begin{table*}[h]
\caption{The frequencies  of the massive particles around the black holes in the unit of $sec^{-1}$.} 
\begin{center}
%-------------------------------
\begin{tabular}{cccccc}
\hline 
  & $a/M$ & $M/M_{\odot}$ & $\omega_{\phi}$ & $\omega_{r}$ & $\omega_{p}$\\
\hline 
GRO~J1655-40 & $~0.65-0.75~$ & $~~6.30~$ & $~319.68-322.86~$ & $283.65-285.08~$ & $~36.03-37.78~$\\
4U~1543-47   & $~0.75-0.85~$ & $~~9.40~$ & $~212.13-214.26~$ & $189.15-190.11~$ & $~22.98-24.15~$\\
\hline
\end{tabular}
%\vskip -5.7cm
\end{center}
\end{table*}
%--------------------------------

%------------ Table 2 --------------------------------------------
\begin{table*}
%\begin{table*}[h]
\caption{The frequencies $\omega_{\phi}=\omega_{r}$ of the massive particles around the black holes in the unit of $sec^{-1}$ in the limit of $r_{s}\gg M$.} 
\begin{center}
%-------------------------------
\begin{tabular}{ccc}
\hline 
  & $M/M_{\odot}$ & $\omega_{\phi}$\\
\hline 
GRO~J1655-40 & $~~6.30~$ & $~3.14\times 10^{-14}~$\\
4U~1543-47   & $~~9.40~$ & $~3.83\times 10^{-14}~$\\
SN~1979C     & $~~(5-10)~$ & $~(2.80-3.95)\times 10^{-14}~$\\
SN~1987A     & $~~20~$ & $~5.60\times 10^{-14}~$\\
SN~2213      & $~~(100-250)~$ & $~(12.50-19.76)\times 10^{-14}~$\\
M87          & $~~2\times 10^{9}$ & $~5.59\times 10^{-10}~$\\
NGC 3115     & $~~1\times 10^{9}$ & $~3.95\times 10^{-10}~$\\
NGC 4594     & $~~5\times 10^{8}$ & $~2.80\times 10^{-10}~$\\
NGC 3377     & $~~1\times 10^{9}$ & $~3.95\times 10^{-10}~$\\
NGC 4258     & $~~4\times 10^{7}$ & $~7.91\times 10^{-11}~$\\
M31          & $~~3\times 10^{7}$ & $~6.85\times 10^{-11}~$\\
M32          & $~~3\times 10^{6}$ & $~2.17\times 10^{-11}~$\\
Galatic center & $~~2.5\times 10^{6}$ & $~1.98\times 10^{-11}~$\\
\hline
\end{tabular}
%\vskip -5.7cm
\end{center}
\end{table*}
%--------------------------------

In the limit of $r_{s}\gg M$, from (\ref{omegaphi}) and (\ref{omegar}) one can obtain the closed bound
orbit with 
\beq
\omega_{\phi}=\omega_{r}=\frac{M^{1/2}}{r_{s}^{3/2}},
\label{omegaphir0}
\eeq
to yield the vanishing $\omega_{p}$.  For a specific example of the massive particles with $\epsilon_{c}=1.0$,
in the stable circular orbit of radius $r_{s}=1.0~ly=9.46\times 10^{12}~km$ around the Kerr
black hole of mass $M$, one can approximately estimate the physical quantities such as 
$l_{c}=1.120\times 10^{12}~(M/M_{\odot})^{1/2}~km^{2}~sec^{-1}$
and $\omega_{\phi}=1.25\times 10^{-14}~(M/M_{\odot})^{1/2}~sec^{-1}$. 
In the exemplary black holes considered above, we then obtain the frequency values 
$\omega_{\phi}=3.14\times 10^{-14}~sec^{-1}$, and $3.83\times 10^{-14}~sec^{-1}$, 
for GRO~J1655-40 and for 4U~1543-47, respectively. Next, we consider the supernovas SN 1979C, SN 1987A and SN 2213-1745.
For the general relativity treatments, we again study this astrophysical problem 
in the framework of the Kerr curved spacetime. For the case of $r_{s}=1.0~ly$, we obtain the explicit frequencies 
$\omega_{\phi}=(2.80-3.95)\times 10^{-14}~sec^{-1},~5.60\times 10^{-14}~sec^{-1},~(12.50-19.76)\times 10^{-14}~sec^{-1}$, 
for the cases of SN 1979C, SN 1987A and for SN~2213, with given masses
(5$-$10) $M_{\odot}$~\cite{nasa}, 20 $M_{\odot}$~\cite{arnett89} and (100$-$250) $M_{\odot}$~\cite{cooke12}, respectively.
Here, we have used the progenitor masses for the supernova masses, for the sake of estimation of the corresponding 
frequencies of the supernovas. 

For examples of the massive particles with $\epsilon_{c}=1.0$,
in the stable circular orbit of radius $r_{s}=1.0~ly=9.46\times 10^{12}~km$ around the supermassive Schwarzschild black 
holes, we also use the formulas in (\ref{omegaphir0}) to yield the frequency values $\omega_{\phi}$ and 
$\omega_{r}$, since the formulas in (\ref{omegaphir0}) are approximately inert to the angular momentum $a$ of the black hole in the above limiting cases. Exploiting the specific data for the potential supermassive Schwarzschild black holes~\cite{hobson06}, we obtain 
the explicit frequencies, $\omega_{\phi}=5.59\times 10^{-10}~sec^{-1}$ for M87 with 2$\times 10^{9}$ $M_{\odot}$, $\omega_{\phi}=3.95\times 10^{-10}~sec^{-1}$ for NGC 3115 with 1$\times 10^{9}$ $M_{\odot}$, $\omega_{\phi}=2.80\times 10^{-10}~sec^{-1}$ for NGC 4594 with 5$\times 10^{8}$ $M_{\odot}$, $\omega_{\phi}=3.95\times 10^{-10}~sec^{-1}$ for NGC 3377 with 1$\times 10^{9}$ $M_{\odot}$, $\omega_{\phi}=7.91\times 10^{-11}~sec^{-1}$ for NGC 4258 with 4$\times 10^{7}$ $M_{\odot}$, $\omega_{\phi}=6.85\times 10^{-11}~sec^{-1}$ for M31 with 3$\times 10^{7}$ $M_{\odot}$, $\omega_{\phi}=2.17\times 10^{-11}~sec^{-1}$ for M32 with 3$\times 10^{6}$ $M_{\odot}$ and $\omega_{\phi}=1.98\times 10^{-11}~sec^{-1}$ for Galatic center with 2.5$\times 10^{6}$ $M_{\odot}$, respectively. Those frequencies $\omega_{\phi}$ for the potential supermassive Schwarzschild black holes are listed in Thable II, together with frequencies $\omega_{\phi}$ for the above rotating black holes.

One can next readily evaluate the velocity of the particle as $v=118.38~(M/M_{\odot})^{1/2}~m~sec^{-1}$ to conclude that the particles in the circular orbit of radius $r_{s}=1.0~ly$ around the black hole of mass up to
$M=7.83~M_{\odot}$ cannot have the velocity greater than that of the shock wave along the azimuthal direction. 
The frequencies of the massive particles around the stable circular orbit of the
black hole play a major role in producing the circular ring with a sensible luminosity around the
black hole. If the massive particles in the perfect fluid described in the previous
section happen to be trapped in the circular orbit at $r=r_{s}$
with $\epsilon_{c}$ and $l_{c}$, (\ref{prho2}) and (\ref{euler2}) become
\beq
\frac{(P+\rho)^{2}}{n^{2}}\frac{1-2Mr_{s}^{-1}+a^{2}r_{s}^{-2}}
{1+a^{2}r_{s}^{-2}+2Ma^{2}r_{s}^{-3}}=\frac{(P_{\infty}
+\rho_{\infty})^{2}}{n_{\infty}^{2}},\label{hydrodynamics22}
\eeq
\beq
\frac{M}{r_{s}^{2}}+\frac{Ma^{2}(r_{s}^{2}+a^{2}-6Mr_{s})}{r_{s}^{4}
(r_{s}^{2}+a^{2}+2Ma^{2}r_{s}^{-1})}+\frac{1-2Mr_{s}^{-1}+a^{2}r_{s}^{-2}}
{P+\rho}\frac{dP}{dr}=0.\label{hydrodynamics2}
\eeq
These equations govern the hydrodynamics of the trapped particles in the circular orbit, in terms of the 
pressure $P$ and internal energy
density $\rho$ of the fluid of those massive particles.

Next, we study the photons following the null geodesics spiraling toward
the rotating Kerr black holes on the equatorial plane, where the photon potential 
is obtainable from (\ref{veff}) with $\kappa=0$. Exploiting the photon potential and 
the condition $\frac{dV}{dr}(r=r_{0})=0$, we obtain the
radius $r_{0}$ of the unstable orbit on the equatorial
plane for a given $\epsilon$ and $l$ to yield 
\beq
r_{0}=\frac{3M(l-a\epsilon)^{2}}{l^{2}-\epsilon^{2}a^{2}},
\eeq 
which shows that, in the gravitational region, the propagation direction of the light changes
following the above unstable orbit. Exploiting the conditions $\frac{dV}{dr}(r=r_{0})=0$ 
and $V(r=r_{0})=0$, one can readily show that $r_{0}$ satisfies the identity 
\beq
r_{0}^{3/2}-3Mr_{0}^{1/2}+2aM^{1/2}=0,
\eeq
which reduces to the Schwarzschild black hole result $r_{0}=3M$ in the $a=0$ limit~\cite{wald}. 
We reemphasize that, since the orbit of the photon is unstable, we do not have a circular trajectory of the
photon, and instead we have a bending of the light in the strong gravitational field originated from
the rotating Kerr black hole of mass $M$.  In fact, the impact parameter defined by
$l/\epsilon$ is evaluated in our case to produce 
\beq
\frac{l^{2}}{\epsilon^{2}}=a^{2}+3r_{0}^{2},
\eeq 
which, in the vanishing $a$ limit, also reduces to the Schwarzschild black hole result, $l^{2}/\epsilon^{2}=27M^{2}$. 

%%%%%%%%%%%%%%%%%%%%%%%%%%%%%%%%%%%%%%%%%%%%%%%%%%%%%%%%
\section{Geometrical aspects of Kerr spacetime}
\setcounter{equation}{0}
\renewcommand{\theequation}{\arabic{section}.\arabic{equation}}
%%%%%%%%%%%%%%%%%%%%%%%%%%%%%%%%%%%%%%%%%%%%%%%%%%%%%%%%

In this section, we investigate the GEMS structure and the warp products of the Kerr black hole 
on the equatorial plane on $\theta=\pi/2$. After some algebra, for the Kerr black hole in the region
$r\ge r_{-}$ we obtain the (9+3) GEMS structure 
\beq
s^{2}=-(dz^{0})^{2}+(dz^{1})^{2}+(dz^{2})^{2}+(dz^{3})^{2}-(dz^{4})^{2}+(dz^{5})^{2}
+(dz^{6})^{2}+(dz^{7})^{2}+(dz^{8})^{2}+(dz^{9})^{2}+(dz^{10})^{2}-(dz^{11})^{2},
\label{011}
\eeq
with the coordinate transformations 
\bea
z^{0}&=&k^{-1}\left(1-\frac{2M}{r}+\frac{a^{2}}{r^{2}}\right)^{1/2}\sinh k t,\nn\\
z^{1}&=&k^{-1}\left(1-\frac{2M}{r}+\frac{a^{2}}{r^{2}}\right)^{1/2}\cosh k t,\nn\\
z^{2}&=&r\cos \phi,\nn\\
z^{3}&=&r\sin \phi,\nn\\
z^{4}&=&k^{-1}\left(\frac{2M}{r}\right)^{1/2}\sinh k (t+a\phi),\nn\\
z^{5}&=&k^{-1}\left(\frac{2M}{r}\right)^{1/2}\cosh k (t+a\phi),\nn\\
z^{6}&=&\left(a^{2}+\frac{4Ma^{2}}{r}\right)^{1/2}\cos\phi,\nn\\
z^{7}&=&\left(a^{2}+\frac{4Ma^{2}}{r}\right)^{1/2}\sin\phi,\nn\\
z^{8}&=&k^{-1}\left(\frac{2M}{r}+\frac{a^{2}}{r^{2}}\right)^{1/2}\cos k t,\nn\\
z^{9}&=&k^{-1}\left(\frac{2M}{r}+\frac{a^{2}}{r^{2}}\right)^{1/2}\sin k t,\nn\\
z^{10}&=&\int dr\left(\frac{(r_{+}+r_{-})r^{2}+r_{+}^{2}(r+r_{+})}{r^{2}(r-r_{-})}
\right)^{1/2},\nn\\
z^{11}&=&\int dr f(r),
\label{gems11} 
\eea
where 
\bea
f(r)&=&\frac{4r_{+}^{5}r_{-}+r_{+}^{4}(r_{+}+r_{-})r}{r^{4}(r_{+}-r_{-})^{2}}
+\frac{(r_{+}+r_{-})^{2}r_{+}r_{-}}{r^{3}[r+2(r_{+}+r_{-})]}+\frac{r_{+}^{4}[(r_{+}+r_{-})r+2r_{+}r_{-}]}
{r^{4}(r_{+}-r_{-})^{2}[(r_{+}+r_{-})r+r_{+}r_{-}]},\nn\\
r_{\pm}&=&M\pm (M^{2}-a^{2})^{1/2},\nn\\ 
k&=&\frac{r_{+}-r_{-}}{2r_{+}^{2}}.
\label{frk}
\eea 
In the region $r\le r_{-}$ we obtain 
the (9+3) GEMS structure in (\ref{011}), 
with the modified coordinate transformations 
where $(z^{0},z^{1},z^{2},z^{3},z^{4},z^{5},z^{6},z^{7},z^{8},z^{9},z^{11})$ are the same as those in (\ref{gems11})
and 
\beq 
z^{10}=\int dr\left(\frac{(r_{+}+r_{-})r^{2}+r_{+}^{2}(r+r_{+})}{r^{2}(r_{-}-r)}\right)^{1/2}.
\eeq

Next, we study the warp products~\cite{bishop69,beem96,choi00,hong05}
of the Kerr black hole on the equatorial plane on $\theta=\pi/2$. To do this, we rewrite the metric (\ref{kerrmetric}) as follows
\beq
ds^{2}=-e^{2\psi_{1}}dt^{2}+e^{2\psi_{2}}(d\phi-\Omega dt)^{2}
+e^{2\psi_{3}}dr+e^{2\psi_{4}}d\theta^{2},
\label{kerrmetric2}
\eeq
where
\beq
e^{2\psi_{1}}=\frac{\Sigma\Delta}{A},~~~e^{2\psi_{2}}=\frac{A\sin^{2}\theta}{\Sigma},~~~
e^{2\psi_{3}}=\frac{\Sigma}{\Delta},~~~e^{2\psi_{4}}=\Sigma.
\eeq
For a comoving observer on the equatorial plane, we find
\beq
ds^{2}=-d\mu^{2}+f_{1}^{2}(\mu) dt^{2}+f_{2}^{2}(\mu) d\phi^{2}.
\label{kerrmetric3}
\eeq 
Here we obtain
\beq
d\mu^{2}=\frac{r^{2}}{(r_{+}-r)(r-r_{-})}dr^{2}
\label{dmu2}
\eeq
which is integrated to yield
\beq
\mu=2r_{-}\cos^{-1}\left(\frac{r_{+}-r}{r_{+}-r_{-}}\right)^{1/2}\equiv F(r),
\eeq
and $f_{1}(\mu)$ and $f_{2}(\mu)$ are given by
\bea
f_{1}(\mu)&=&\left(\frac{(F^{-1}(\mu))^{2}(r_{+}-F^{-1}(\mu))(F^{-1}(\mu)-r_{-})}
{((F^{-1}(\mu))^{2}+a^{2})^{2}+a^{2}(r_{+}-F^{-1}(\mu))(F^{-1}(\mu)-r_{-})}\right)^{1/2} \nn\\
f_{2}(\mu)&=&\left(\frac{((F^{-1}(\mu))^{2}+a^{2})^{2}+a^{2}(r_{+}-F^{-1}(\mu))(F^{-1}(\mu)-r_{-})}
{(F^{-1}(\mu))^{2}}\right)^{1/2}.
\eea

We then find the Ricci tensors as follows
\bea
R_{\mu\mu}&=&-\frac{f_{1}^{\pr\pr}}{f_{1}}-\frac{f_{2}^{\pr\pr}}{f_{2}},\nn\\
R_{tt}&=&f_{1}f_{1}^{\pr\pr}+\frac{f_{1}f_{1}^{\pr}f_{2}^{\pr}}{f_{2}},\nn\\
R_{\phi\phi}&=&f_{2}f_{2}^{\pr\pr}+\frac{f_{2}f_{2}^{\pr}f_{1}^{\pr}}{f_{1}},
\eea
to yield the Einstein scalar
\beq
R=2\left(\frac{f_{1}^{\pr\pr}}{f_{1}}+\frac{f_{2}^{\pr\pr}}{f_{2}}
+\frac{f_{1}^{\pr}f_{2}^{\pr}}{f_{1}f_{2}}\right).
\eeq
Here the primes denote the derivatives with respect to $\mu$.

%%%%%%%%%%%%%%%%%%%%%%%%%%%%%%%%%%%%%%%%%%%%%%%%%%%%%%%%
\section{Conclusions}
\setcounter{equation}{0}
\renewcommand{\theequation}{\arabic{section}.\arabic{equation}}
%%%%%%%%%%%%%%%%%%%%%%%%%%%%%%%%%%%%%%%%%%%%%%%%%%%%%%%%

On the equatorial plane of the Kerr black hole, we have constructed
the global flat embedding structures possessing (3+9) dimensionalities outside and 
inside the event horizon of the rotating Kerr black hole. Moreover, on this plane we have investigated the warp 
products of the Kerr spacetime.

We have introduced the rotating Kerr black hole, to investigate hydrodynamics of the massive particles and
massless photons. We have constructed relations among number density, pressure and internal energy
density of the massive particles and photons around the rotating Kerr black hole to figure out an accretion onto 
the black hole. On the equatorial plane of the Kerr black hole, we have investigated the bound orbits of the massive 
particles and photons around the black hole to produce their radial, azimuthal and precession 
frequencies. With these frequencies we have studied the black holes GRO~J1655-40 and 4U~1543-47, 
to explicitly evaluate the radial, azimuthal and precession frequencies of the massive particles in the 
flow of the perfect fluid. 

We have next considered the massive particles in the stable circular orbit of radius of $1.0~ly$ around these
black holes, around the supernovas SN 1979C, SN 1987A and SN 2213-1745 
in the Kerr curved spacetime, and around the potential supermassive Schwarzschild black holes M87, NGC 3115, NGC 4594, 
NGC 3377, NGC 4258, M31, M32 and Galatic center, to estimate their radial and azimuthal frequencies, 
which are shown to be the same resulting in no precession motion. The photon unstable orbit has been also 
discussed in terms of the impact parameter of the photon trajectory. 

Finally, we have comments on the relation of hydrodynamics of the massive/massless particles and the global 
flat embedding structures of the rotating Kerr black hole. One of the features of solutions to the field equations 
of general relativity is the presence of singularities. In order to remove the coordinate singularities, 
the higher dimensional flat embedding of the black hole solution has been exploited. For instance, for the Schwarzschild black hole, the 
(5+1)-dimensional flat embedding was constructed~\cite{deser97,deser972,deser99} to remove the coordinate singularity and to 
study a thermal Hawking effect on the curved manifold~\cite{hawk75,hawk752} related with the Unruh effect~\cite{unr} in the 
higher dimensional manifold. Recently, a free fall temperature of the Schwarzschild-AdS black hole was evaluated in 
the higher dimensional flat spacetime~\cite{thorlacius08}. The free fall temperature for the Gibbons-Maeda-Garfinkle-Horowitz-Strominger 
spacetime~\cite{gibbons88,gibbons882} was also evaluated~\cite{park14}. The thermodynamics of the massive/massless particles is thus closely 
related to the  global flat embedding structures of the rotating Kerr black hole. In this work, the relation of hydrodynamics 
of the massive/massless particles and their thermodynamics associated with the global flat embedding structures of the rotating 
Kerr black hole is not explicitly described. It will be interesting to study further the hydrodynamics and 
thermodynamics of the massive/massless particles by including thermal aspects of the particles in their hydrodynamics 
which is investigated in this work.

\end{document}